\documentclass[aps,pra,twocolumn,groupedaddress]{revtex4-1}

\usepackage{graphicx}  
\usepackage{dcolumn}   
\usepackage{bm}        
\usepackage{amssymb}   
\usepackage{pdfpages}
\usepackage{hyperref}

\usepackage{mathrsfs}
\usepackage{amsmath}

\usepackage{graphicx}

\begin{document} 

\title{Quantum thermalization through entanglement in an isolated many-body system}

\author{A. M.~Kaufman}
\author{M. E.~Tai}
\author{A. Lukin}
\author{M.~Rispoli}
\author{R.~Schittko}
\author{P. M.~Preiss}
\author{M.~Greiner}
\email[E-mail: ]{greiner@physics.harvard.edu}
\address{Department of Physics, Harvard University, Cambridge, Massachusetts 02138, USA}

\date{\today}


\begin{abstract}

The concept of entropy is fundamental to thermalization, yet appears at odds with basic principles in quantum mechanics. Statistical mechanics relies on the maximization of entropy for a system at thermal equilibrium. However, an isolated many-body system initialized in a pure state will remain pure during Schr\"{o}dinger evolution, and in this sense has static, zero entropy. The underlying role of quantum mechanics in many-body physics is then seemingly antithetical to the success of statistical mechanics in a large variety of systems. Here we experimentally study the emergence of statistical mechanics in a quantum state, and observe the fundamental role of quantum entanglement in facilitating this emergence. We perform microscopy on an evolving quantum system, and we see thermalization occur on a local scale, while we measure that the full quantum state remains pure. We directly measure entanglement entropy and observe how it assumes the role of the thermal entropy in thermalization. Although the full state remains measurably pure, entanglement creates local entropy that validates the use of statistical physics for local observables. In combination with number-resolved, single-site imaging, we demonstrate how our measurements of a pure quantum state agree with the Eigenstate Thermalization Hypothesis and thermal ensembles in the presence of a near-volume law in the entanglement entropy. 

\end{abstract}
\maketitle 

\begin{figure}[h!]
	\centering
	\includegraphics[scale=1.35]{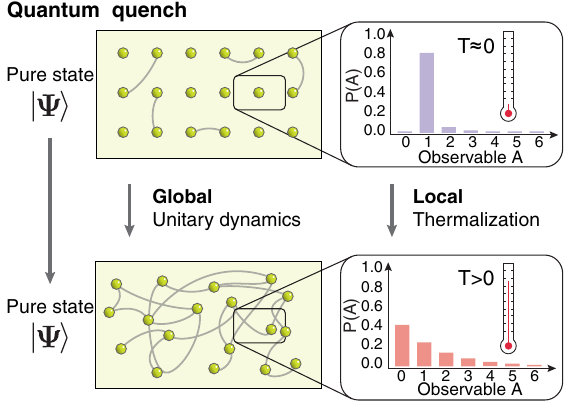}
	\caption{{\bf Schematic of thermalization dynamics in closed systems}.  An isolated quantum system at zero temperature can be described by a single pure wavefunction $\vert \Psi \rangle$. Subsystems of the full quantum state appear pure, as long as the entanglement (indicated by grey lines) between subsystems is negligible. If suddenly perturbed, the full system evolves unitarily, developing significant entanglement between all parts of the system. While the full system remains in a pure, and in this sense zero-entropy state, the entropy of entanglement causes the subsystems to equilibrate, and local, thermal mixed states appear to emerge within a globally pure quantum state.}
	\label{fig:conceptual}
\end{figure}

When an isolated quantum system is significantly perturbed, for instance due to a sudden change in the Hamiltonian, we can predict the ensuing dynamics with the resulting eigenstate distribution induced by the perturbation or so-called ``quench"~\cite{Sakurai}. At any given time, the evolving quantum state will have amplitudes that depend on the eigenstates populated by the quench, and the energy eigenvalues of the Hamiltonian. In many cases, however, such a system can be extremely difficult to simulate, often because the resulting dynamics entail a large amount of entanglement~\cite{CardyOneD, Amico2008,Daley2012, Schachenmayer2013}. Yet, surprisingly, this same isolated quantum system can thermalize under its own dynamics unaided by a reservoir (Figure~\ref{fig:conceptual})~\cite{Deutsch1991, Olshanii2008,Eisert2015}, so that the tools of statistical mechanics apply and challenging simulations are no longer required. In this case, a quantum state coherently evolving according to the Schr\"{o}dinger equation is such that most observables can be predicted from a thermal ensemble and thermodynamic quantities. Strikingly, even with infinitely many copies of this quantum state, these same observables are fundamentally unable to reveal whether this is a single quantum state or a thermal ensemble. In other words, a globally-pure quantum state is apparently indistinguishable from a mixed, globally-entropic thermal ensemble~\cite{Shankar1985, Deutsch1991, SrendickiETH, Olshanii2008}. Ostensibly the coherent quantum amplitudes that define the quantum state in Hilbert space are no longer relevant, even though they evolve in time and determine the expectation values of observables. The dynamic convergence of the measurements of a pure quantum state to the predictions of a thermal ensemble, and the physical process by which this convergence occurs, is the experimental focus of this work. 

On-going theoretical studies over the past three decades~\cite{Shankar1985, Deutsch1991, SrendickiETH, Olshanii2008, Rigol2012, Deutsch2013,Alessio2015} have, in many regards, clarified the role of quantum mechanics in statistical physics. The conundrum surrounding the agreement of pure states with extensively entropic thermal states is resolved by the counter-intuitive effects of quantum entanglement. A canonical example of this point is the Bell state of two spatially separated spins: while the full quantum state is pure, local measurements of just one of the spins reveals a statistical mixture of reduced purity. This local statistical mixture is distinct from a superposition, because no operation on the single spin can remove these fluctuations or restore its quantum purity. In such a way, the spin's entanglement with another spin creates local entropy, called entanglement entropy. Entanglement entropy is not a phenomenon restricted to spins, but exists in all quantum systems that exhibit entanglement. And while probing entanglement is a notoriously difficult experimental problem, this loss of local purity, or, equivalently, the development of local entropy, establishes the presence of entanglement when it can be shown that the full quantum state is pure. 

In this work, we directly observe a globally pure quantum state dynamically lose local purity to entanglement, and in parallel become locally thermal. Recent experiments have demonstrated analogies between classical chaotic dynamics and the role of entanglement in few-qubit spin systems~\cite{Martinis2016}, as well as the dynamics of thermalization within an ion system~\cite{Schaetz2015}. Furthermore, studies of bulk gases have shown the emergence of thermal ensembles and the effects of conserved quantities in isolated quantum systems through macroscopic observables and correlation functions~\cite{Trotzky2012,Langen2013,Geiger2014,Langen2015}. We are able to directly measure the global purity as thermalization occurs through single-particle resolved quantum many-body interference. In turn, we can observe microscopically the role of entanglement in producing local entropy in a thermalizing system of itinerant particles, which is paradigmatic of the systems studied in statistical mechanics.  

In such studies, we will explore the equivalence between the entanglement entropy we measure and the expected thermal entropy of an ensemble~\cite{Rigol2012, Deutsch2013}. We further address how this equivalence is linked to the Eigenstate Thermalization Hypothesis (ETH), which provides an explanation for thermalization in closed quantum systems~\cite{Shankar1985, Deutsch1991, SrendickiETH, Olshanii2008}. ETH is typically framed in terms of the small variation of observables (expectation values) associated with eigenstates close in energy~\cite{Deutsch1991, SrendickiETH, Olshanii2008}, but the role of entanglement in these eigenstates is paramount~\cite{Deutsch2013}. Indeed, fundamentally, ETH implies an equivalence of the local reduced density matrix of a single excited energy eigenstate and the local reduced density matrix of a globally thermal state~\cite{Huse2015}, an equivalence which is made possible only by entanglement and the impurity it produces locally within a global pure state. The equivalence between these two seemingly distinct systems, the subsystems of a quantum pure state and a thermal ensemble, ensures thermalization of most observable quantities after a quantum quench. Through parallel measurements of the entanglement entropy and local observables within a many-body Bose-Hubbard system, we are able to experimentally study this equivalence at the heart of quantum thermalization. 

\section*{Experimental protocol}

\begin{figure}[h!]
	\centering
	\includegraphics[scale=.75]{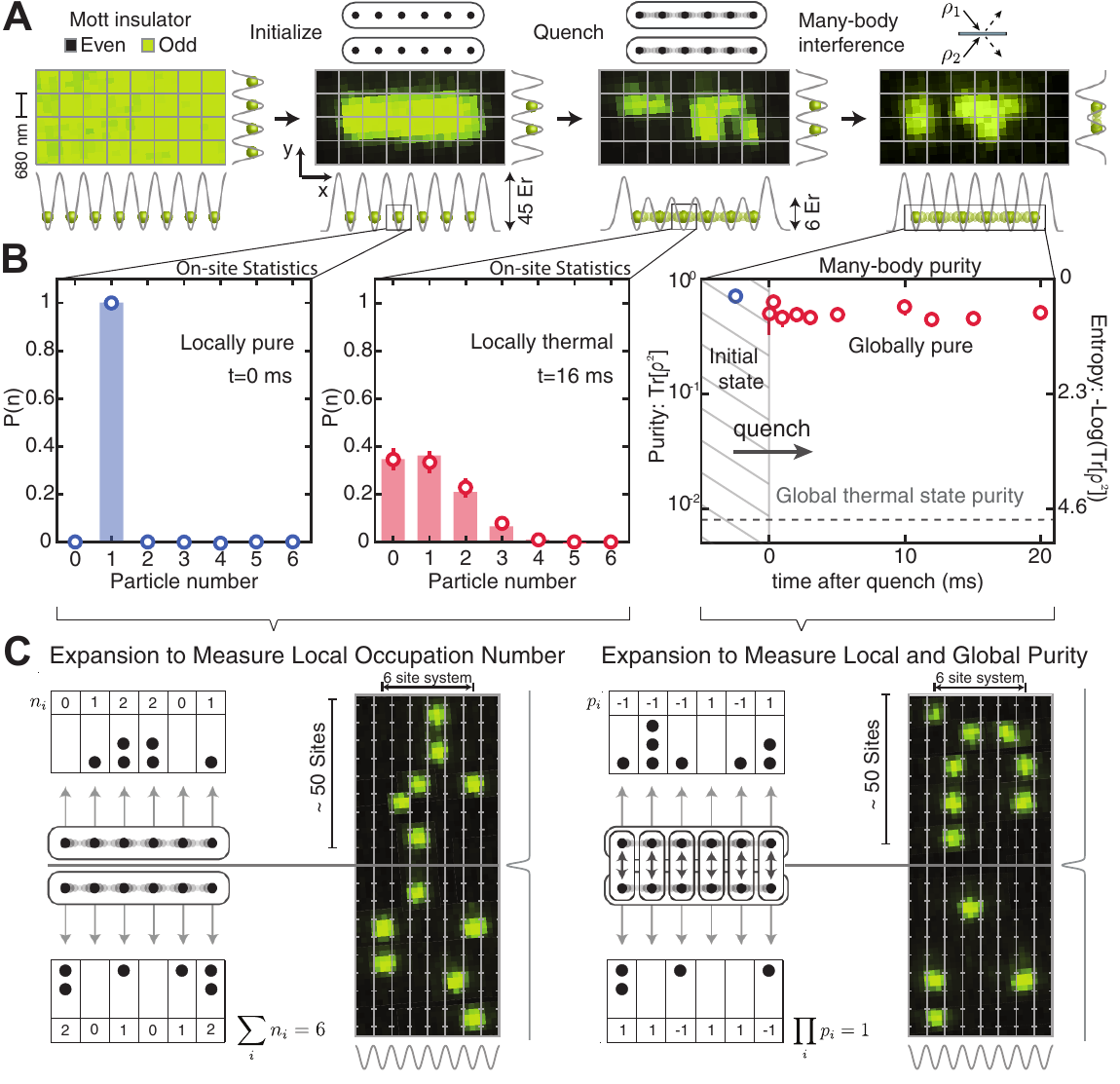}
	\caption{{\bf Experimental sequence} {\bf (A)} Using tailored optical potentials superimposed on an optical lattice, we deterministically prepare two copies of a six-site Bose-Hubbard system, where each lattice site is initialized with a single atom. We enable tunneling in the x-direction and obtain either the ground state (adiabatic melt) or a highly excited state (sudden quench) in each six-site copy. After a variable evolution time, we freeze the evolution and characterize the final quantum state by either acquiring number statistics or the local and global purity. {\bf (B)} We show site-resolved number statistics of the initial distribution (first panel, strongly peaked about one atom with vanishing fluctuations), or at later times (second panel) to which we compare the predictions of a canonical thermal ensemble of the same average energy as the quenched quantum state ($J/(2\pi) = 66~\mathrm{Hz}, U/(2\pi) = 103~\mathrm{Hz})$. Alternatively, we can measure the global many-body purity, and observe a static, high purity. This is in stark contrast to the vanishing global purity of the canonical thermal ensemble, yet this same ensemble accurately describes the local number distribution we observe. {\bf (C)} To measure the atom number locally, we allow the atoms to expand in half-tubes along the $y$-direction, while pinning the atoms along $x$. In separate experiments, we apply a many-body beam splitter by allowing the atoms in each column to tunnel in a projected double-well potential. The resulting atom number parity, even or odd, on each site encodes the global and local purity.
	}

	\label{fig:protocol}
\end{figure} 

For our experiments, we utilize a Bose-Einstein condensate of $^{87}$Rb atoms
loaded into a two-dimensional optical lattice that lies at the focus of a high
resolution imaging system~\cite{GreinerQGM, BlochQGM}. The system is described by the Bose-Hubbard
Hamiltonian,
\begin{multline}
\label{BHM}
 H = - (J_x \sum_{x,y} a^\dagger_{x,y} a_{x+1,y} + J_y \sum_{x,y} a^\dagger_{x,y}
a_{x,y+1} +h.c.) \\ +\frac{U}2\sum_{x,y} n_{x,y}(n_{x,y} - 1),
\end{multline}
where $a_{x,y}^\dagger$, $a_{x,y}$, and $n_{x,y} = a_{x,y}^\dagger a_{x,y}$ are the bosonic
creation, annihilation, and number operators at the site located at $\{x,y\}$, respectively. Atoms
can tunnel between neighboring lattice sites at a rate $J_{i}$ and experience a
pairwise interaction energy $U$ when multiple atoms occupy a site. We have
independent control over the tunneling amplitudes $J_x$ and $J_y$ through the
lattice depth, which can be tuned to yield $J/U \ll 1$ to $J/U \gg 1$. In
addition to the optical lattice, we are able to superimpose arbitrary
potentials using a digital micromirror device (DMD) placed in the Fourier
plane of our imaging system~\cite{Zupancic2016}.

\begin{figure*}[t!]
	\centering
	\includegraphics[scale=1.1]{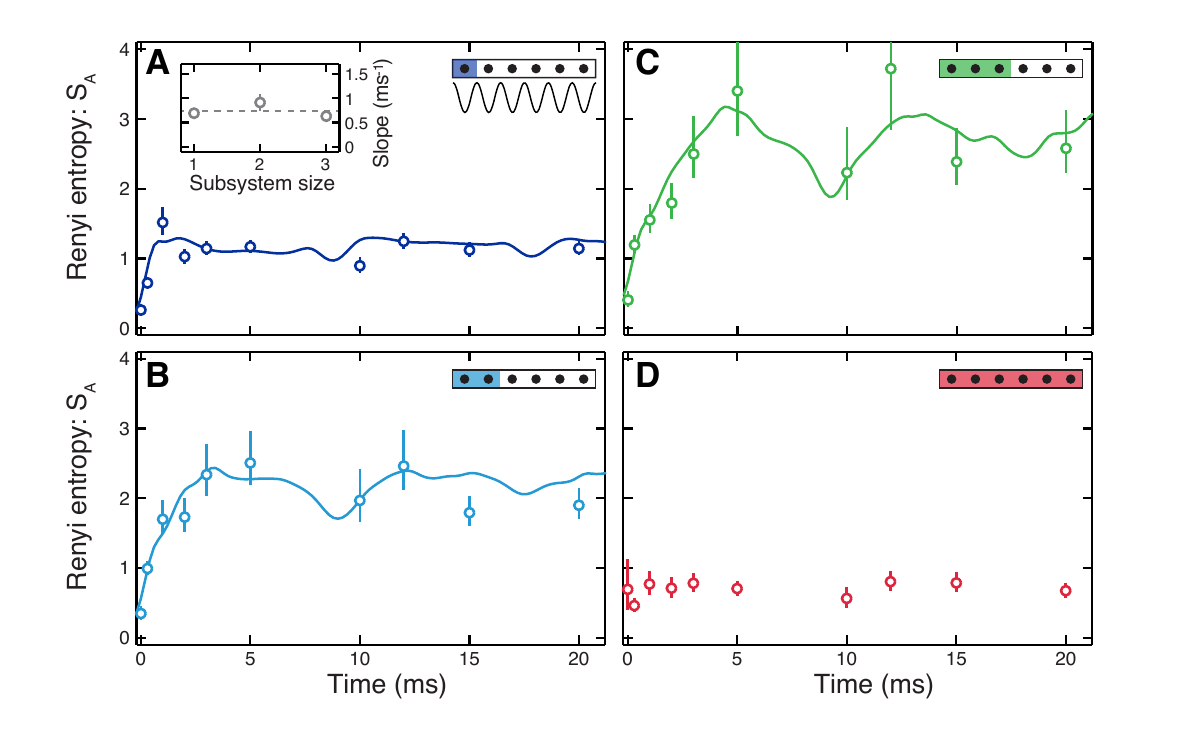}
	\caption{{\bf Dynamics of entanglement entropy. } Starting from a low-entanglement ground state, a global quantum quench leads to the development of large-scale entanglement between all subsystems. We quench a six-site system from the Mott insulating product state ($J/U\ll 1$) with one atom per site to the weakly interacting regime of $J/U=0.64$ ($J/(2\pi) = 66~\mathrm{Hz}$) and measure the dynamics of the entanglement entropy. As it equilibrates, the system acquires local entropy while the full system entropy remains constant and at a value given by measurement imperfections. The dynamics agree with exact numerical simulations with no free parameters (solid lines). Error bars are the standard error of the mean (S.E.M.). For the largest entropies encountered in the three-site system, the large number of populated microstates leads to a significant statistical uncertainty in the entropy, which is reflected in the upper error bar extending to large entropies or being unbounded. Further information about statistics is discussed in the Supplementary Materials~\cite{Supplement}. Inset: slope of the early time dynamics, extracted with a piecewise linear fit~\cite{Supplement}. The dashed line is the mean of these measurements.}
	\label{fig:EEDyn}
\end{figure*} 

To initiate the experiment, we isolate a $2 \times
6$ plaquette from a larger low-entropy Mott insulator with unity filling as
shown in Figure~\ref{fig:protocol}A~\cite{Supplement}. At this point, each system is in a product state of single-atom Fock
states on each of the constituent sites. We then suddenly switch on tunneling in the $x$-direction while the $y$-direction tunneling is suppressed. Each chain is restricted to the original six sites by introducing a barrier at the ends of the chains to prevent tunneling out of the system. These combined steps quench the six-site chains into a Hamiltonian for which the initial state represents a highly excited state that has significant overlap with an appreciable number of energy eigenstates. Each chain represents an identical but independent copy of a quenched system of six particles on six sites, which evolves in the quenched Hamiltonian for a controllable duration. 

In the data that follow, we realize measurements of the quantum purity
and on-site number statistics (Figure~\ref{fig:protocol}C). For measurements of the former, we append to
the quench evolution a beam splitter operation that interferes the two
identical copies by freezing dynamics along the chain and allowing for
tunneling in a projected double-well potential for a prescribed time~\cite{Islam2015}. In the last step for both measurements, a potential
barrier is raised between the two copies and a one-dimensional time-of-flight
in the direction transverse to the chain is performed to measure the resulting
occupation on each site of each copy. 

The ability to measure quantum purity is crucial to assessing the role of entanglement in our system. Tomography of the full quantum state would typically be required to extract the global purity, which is particularly challenging in the full 462-dimensional Hilbert space defined by the itinerant particles in our system. Furthermore, while in spin systems global rotations can be employed for tomography~\cite{Sackett2000}, there is no known analogous scheme for extracting the full density matrix of a many-body state of itinerant particles. The many-body interference described here, however, allows us to extract quantities that are quadratic in the density matrix, such as the purity~\cite{Islam2015}. After performing the beam splitter operation, we can obtain the quantum purity of the full system and any subsystem simply by counting the number of atoms on each site of one of the six-site chains (Figure~\ref{fig:protocol}C). Each run of the experiment yields the parity $P^{(k)} = \Pi_i p^{(k)}_i$, where $i$ is iterated over a set of sites of interest in copy-$k$. The single-site parity operator $p^{(k)}_i$ returns 1 (-1) when the atom number on site-$i$ is even (odd). It has been shown that the beam splitter operation yields $\langle P^{(1)} \rangle = \langle P^{(2)} \rangle = \mathrm{Tr}\left (\rho_1 \rho_2  \right)$, where $\rho_i$ is the density matrix on the set of sites considered for each copy~\cite{JakschPRA, Daley2012,Islam2015}. Because the preparation and quench dynamics for each copy are identical, yielding $\rho_1 = \rho_2 \equiv \rho$, the average parity reduces to the purity: $\langle P^{(k)} \rangle = \mathrm{Tr}(\rho^2)$. When the set of sites considered comprises the full six-site chain, the expectation value of this quantity returns the global many-body purity, while for smaller sets it provides the local purity of the respective subsystem. 

By studying measurements with and without the beam splitter, our data immediately illustrates the contrast between the global and local behavior and how thermalization is manifest (Figure~\ref{fig:protocol}B). We observe that the global many-body state retains its quantum purity in time, affirming the unitarity of its evolution following the quench. This global measurement also clearly distinguishes the quantum state we produce from a canonical thermal ensemble with orders of magnitude smaller purity. Yet, we observe that the number statistics locally converge to a distribution of thermal character, which can be faithfully modeled by that same thermal ensemble. In what follows, we experimentally explore the question suggested by this observation: how does a pure state that appears globally distinct from a thermal ensemble possess local properties that mirror this thermal state? 

The growth of entanglement following a quench is key to understanding how entropy forms within the subsystems of a pure quantum state, thereby facilitating thermalization~\cite{CardyOneD, Daley2012, Schachenmayer2013, Hazzard2014}. When two parts of a system are entangled, the full quantum state $\rho$ cannot be written in a separable fashion with respect to the Hilbert spaces of the subsystems~\cite{HHHH,Horodecki1996a}. As has been shown theoretically~\cite{JakschPRA, Daley2012} and recently observed experimentally~\cite{Islam2015}, this causes the subsystems $\rho_A$ and $\rho_B$ to be in an entropic mixed state even though the full many-body quantum state is pure~\cite{Horodecki1996a}. The mixedness of the subsystem can be quantified by the second-order R\'{e}nyi entropy $S_A = -\textrm{Log}(\textrm{Tr}[\rho_A^2])$, which is the logarithm of the purity of the subsystem density matrix. While the von Neumann entropy is typically used in the context of statistical mechanics, both quantities grow as a subsystem density matrix becomes mixed and increasingly entropic. In the R\'{e}nyi case, the purity in the logarithm quantifies the number of states contributing to the statistical mixture described by the density matrix.  

\section*{Entanglement entropy dynamics and saturation}

We first study the dynamics of the entanglement entropy immediately following the quench, for varying subsystem sizes (Figure~\ref{fig:EEDyn}). Initially, we observe an approximately linear rise in the entropy, with similar slope among the subsystems considered (Figure~\ref{fig:EEDyn} inset)~\cite{CardyOneD}. After an amount of time that depends on the subsystem size, the entanglement entropy saturates to a steady-state value, about which there are small residual temporal fluctuations. The presence of residual fluctuations are, in part, attributable to the finite-size of our system. An exact numerical calculation of the dynamics with no free parameters shows excellent agreement with our experimental measurements. Crucially, the data indicate that while the subsystems acquire entropy in time (Figure~\ref{fig:EEDyn},A-C), the full system entropy remains constant and is small throughout the dynamics (Figure~\ref{fig:EEDyn}D). The high purity of the full system allows us to conclude that the dynamical increase in entropy in the subsystems originates in the propagation of entanglement between the system's constituents. The approximately linear rise at early times (Figure~\ref{fig:EEDyn} inset) is related to the spreading of entanglement in the system within an effective light cone~\cite{CardyOneD,lightcone,Richerme2014}. Furthermore, in analogy to the growth of thermodynamic entropy in an equilibrating classical mechanical system, such as a gas in a closed container, we observe the growth of local entropy in a closed quantum mechanical system. In the quantum mechanical case, however, the mechanism responsible for entropy is entanglement, which is absent from a system modeled by classical mechanics.

When a system thermalizes, we expect that the saturated values of local observables should correspond to the predictions of a statistical ensemble. By analogy, if the entanglement entropy plays the role of thermal entropy, then in a thermalized pure state we expect extensive growth in the entanglement entropy with subsystem volume. When the entanglement entropy in a quantum state grows linearly with the size of the subsystem considered, it is known as a volume law. Ground-breaking theoretical work using conformal field theory has shown that indeed, at long times, a volume law is expected for a quenched, infinite, continuous system, while only an area law with a log correction is expected for the ground state~\cite{CardyGS, CardyOneD, Plenio2010}. Characterizing the large amount of entanglement associated with a volume law is particularly challenging because it results in nearly every entry of the density matrix having small, but importantly non-zero magnitude. 

\begin{figure*}
	\centering
	\includegraphics[scale=.95]{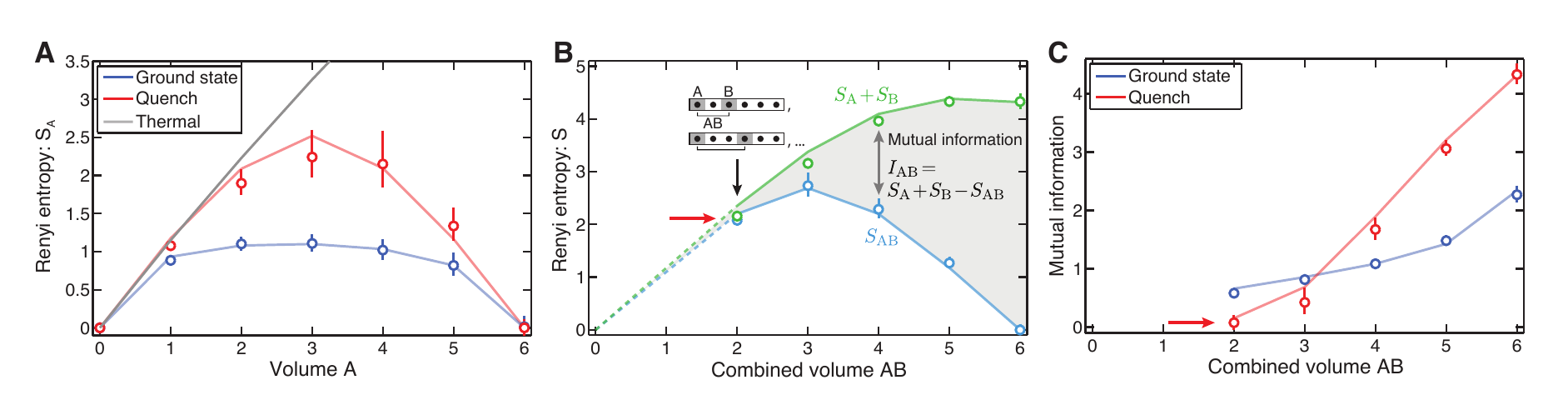}
	\caption{{\bf Thermalized many-body systems.} After the quench, the many-body state reaches a thermalized regime with saturated entanglement entropy. {\bf (A)} In contrast to the ground state, for which the R\'{e}nyi entropy only weakly depends on subsystem size, the entanglement entropy of the saturated, quenched state grows almost linearly with size. As the subsystem size becomes comparable to the full system size, the subsystem entropy bends back to near zero, reflecting the globally pure zero-entropy state. For small subsystems, the R\'{e}nyi entropy in the quenched state is nearly equal to the corresponding thermal entropy from the canonical thermal ensemble density matrix. {\bf (B)} The mutual information $I_{AB} = S_A + S_B - S_{AB}$ quantifies the amount of classical (statistical) and quantum correlations between subsystems A and B. For small subsystems, the thermalized quantum state has $S_A + S_B \approx S_{AB}$ due to the near volume law scaling (red arrow), leading to vanishing mutual information. When the volume of $AB$ approaches the system size, the mutual information will grow because $S_A + S_B$ exceeds $S_{AB}$. {\bf (C)} We study $I_{AB}$ vs the volume of AB for the ground state and the thermalized quenched state. For small system sizes, the quenched state exhibits smaller correlations than the adiabatically prepared ground state, and is nearly vanishing. When probed on a scale near the system size, the highly entangled quenched state exhibits much stronger correlations than the ground state. Throughout this figure, the entanglement entropies from the last time point in Fig. 3 are averaged over all relevant partitionings with the same subsystem volume; we also correct for the extensive entropy unrelated to entanglement~\cite{Supplement}. All solid lines are theory with no free parameters.}
	\label{fig:volume}
\end{figure*}

Using the techniques outlined in this work, we show measurements displaying a near volume law in the entanglement entropy (Figure~\ref{fig:volume}A). A linear growth with volume in the entanglement entropy occurs when each subsystem incoherently populates a number of states that scales with the size of the subsystem Hilbert space. This is because, for the Bose-Hubbard model, the Hilbert space is approximately exponential in the lattice size, which results in a linear growth in $S_A = -\textrm{Log}(\textrm{Tr}[\rho_{A}^2])$. Furthermore, the exact slope of the entanglement entropy versus subsystem volume depends on the average energy of the thermalized pure state~\cite{Grover2015}.  By contrast, we can prepare the ground state of the quenched Hamiltonian by adiabatically reducing the lattice depth. Here, the superfluid ground state of the Bose-Hubbard model has suppressed entanglement, which is predicted to incur slow logarithmic growth in the entanglement entropy with subsystem volume~\cite{CardyGS} . Our measurements clearly distinguish the two cases. The back-bending of the entanglement entropy as the subsystem surpasses half the system size indicates that the state is globally pure. In the quenched state, the high global purity is striking in a state that locally appears completely dephased, which is behavior often associated with environmentally-induced decoherence or other noise sources. 
 
We further observe near quantitative agreement between the exact dependence of the entanglement entropy with subsystem volume and the prediction of a thermal ensemble. We make this comparison by computing a canonical thermal ensemble $\rho^T$ with an average energy that is the same as the quenched quantum state produced experimentally~\cite{Grover2015}. The gray line in Figure~\ref{fig:volume}A is the R\'{e}nyi (thermal) entropy as a function of subsystem size for this calculated thermal state. Although our limited system size prevents comparison over a large range of subsystem sizes, the initial rise of the entanglement entropy with subsystem volume mimics that of the thermal entropy. Despite their similarity, it is worth emphasizing the disparate character of the thermal and entanglement entropy. The entanglement entropy (either the R\'{e}nyi or von Neumann) is instantaneously present in the pure quantum state after coherent unitary evolution, arising from the non-separability of the quantum state between the subsystem and traced out degrees of freedom. On the other hand, the von Neumann thermal entropy within a subsystem of a mixed thermal state is the thermodynamic entropy in statistical mechanics, which could be extracted from irreversible heat flow experiments on the subsystem~\cite{Deutsch2013}. Therefore, the similarity of the R\'{e}nyi  entropies we observe points to an experimental equivalence between the entanglement and thermodynamic entropy~\cite{Grover2014,Grover2015}.

The behavior of the entanglement entropy provides a clean framework for understanding the entropy within thermalizing, closed quantum systems. However, one of the most famous features of entanglement, the presence of non-local correlations, appears inconsistent with what one expects of thermalized systems. In particular, the massive amount of entanglement implied by a volume law suggests a large amount of correlation between disparate parts of the system, while a key feature of a thermal state is the very absence of these long-range correlations. A useful metric for correlations, both classical (statistical) and quantum, between two subsystems $A$ and $B$ is the mutual information $S_A +S_B - S_{AB}$~\cite{Wolf2008, Islam2015}. The mutual information demonstrates that the amount of correlation in the presence of a volume law is vanishing for subsystem volumes that sample less than half the full system, which is where the entropy growth is nearly linear (Figure~\ref{fig:volume}B,C). Furthermore, even though the thermalized quantum state carries more entanglement entropy than the ground state, small subsystems display smaller correlations than in the superfluid ground state. Once the subsystem volume is comparable to the system size, which is where the entanglement entropy deviates from the volume law, the quantum correlations entailed by the purity of the full system become apparent (Figure~\ref{fig:volume}C). The mutual information therefore illustrates how the volume law in the entanglement entropy yields an absence of correlations between sufficiently local observables, even though the quantum state retains a large amount of entanglement. 

 \begin{figure}[t!]
	\centering
	\includegraphics[scale=1]{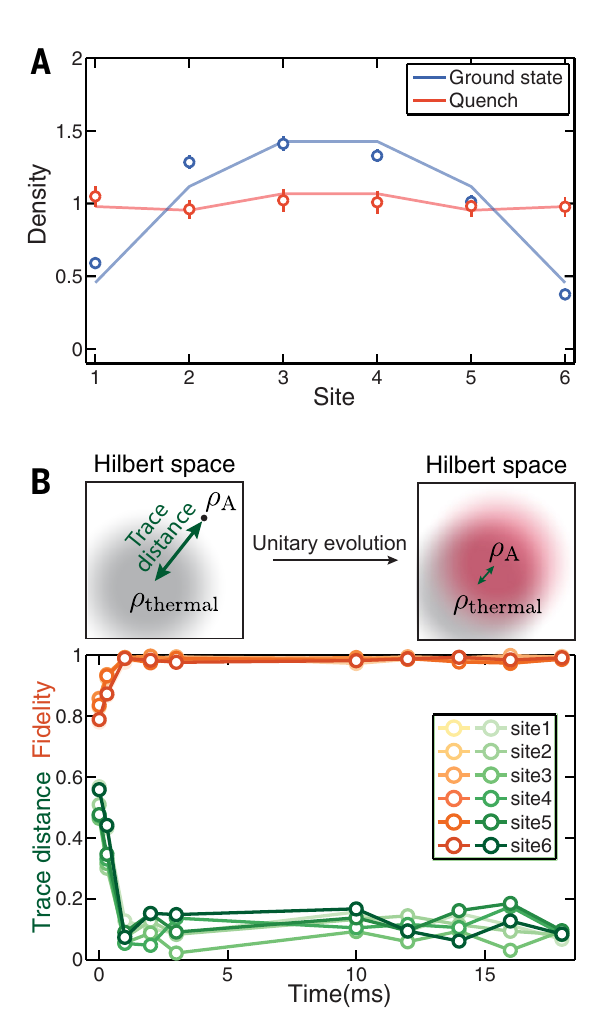}
	\caption{{\bf Observation of local thermalization.} {\bf(A)} After quenching to $J/U =2.6$, the saturated average particle number on each site (density) is nearly equal among the sites of the system, which resembles a system at thermal equilibrium. By comparison, the ground state for the same Bose-Hubbard parameters has significant curvature.  {\bf(B)} In measuring the probabilities to observe a given particle number on a single site, we can obtain the local, single-site density matrix and observe the approach to thermalization. Using two different metrics, we compare the mixed state observed to the mixed state derived from the subsystem of a canonical thermal ensemble, after a quench to $J/U =0.64$. The trace distance provides an effective distance between the mixed states in Hilbert space, while the fidelity is an overlap measure for mixed states. The two metrics illustrate how the pure state subsystem approaches the thermal ensemble subsystem shortly after the quench. The starting value of these quantities is given by the overlap of the initial pure state with the thermal mixed state. Solid lines connect the data points.}
	\label{fig:local}
\end{figure}

\begin{figure*}[t!]
	\centering
	\includegraphics[scale=1]{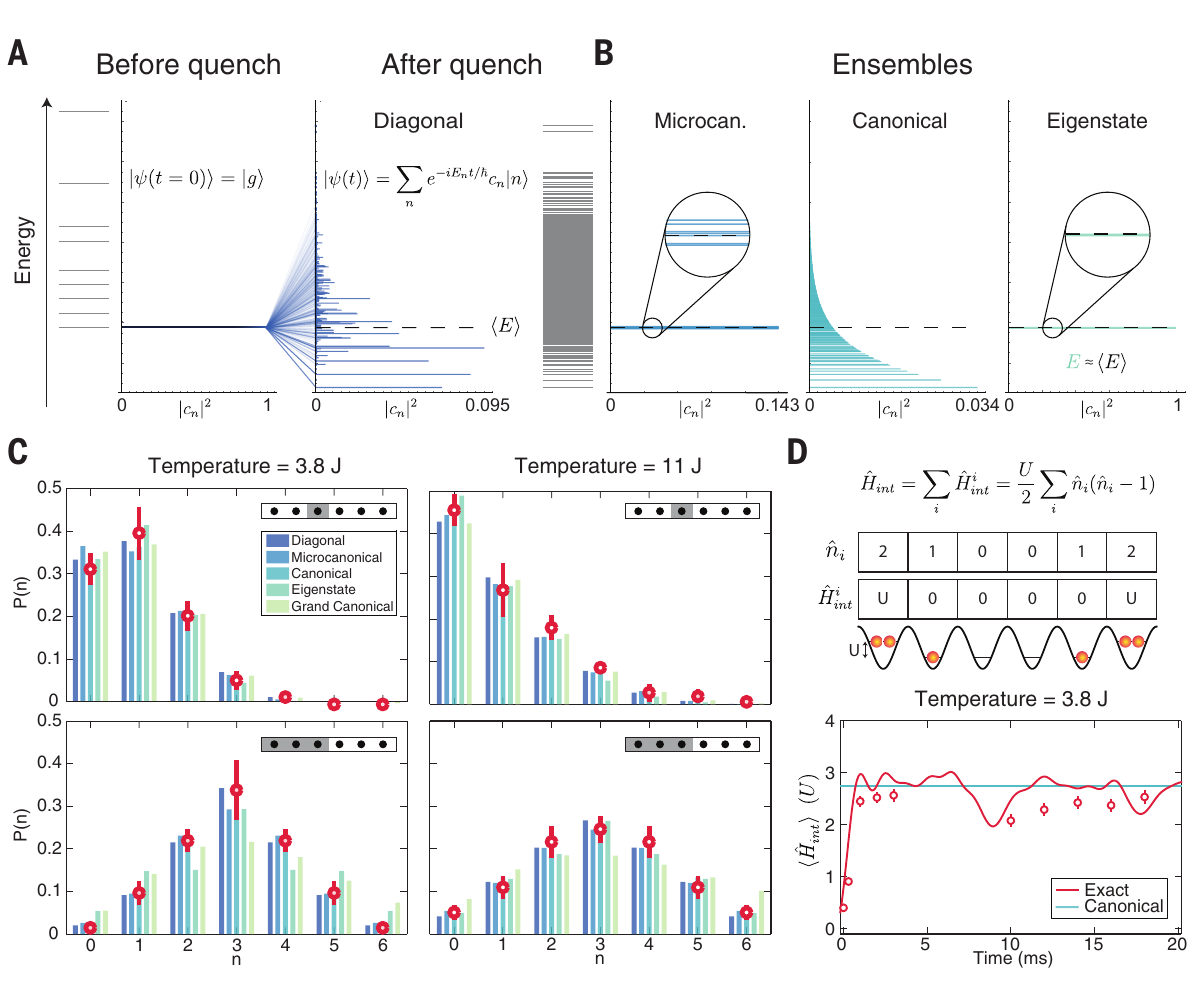}
	\caption{{\bf Local observables in a globally pure quenched state. } {\bf (A)} In a quench, the ground state of the initial Hamiltonian (represented in its eigenbasis in the first panel) is projected onto many eigenstates of the new Hamiltonian and undergoes unitary evolution. According to the ETH, the expectation value of observables at long times can be obtained from a diagonal ensemble (illustrated by the probability weights in the eigenstates of the quenched Hamiltonian) and microcanonical ensemble comprised of such eigenstates. {\bf (B)} Along with the microcanonical ensemble, several other closely related ensembles are compared to the data. {\bf (C)} Thermalization of local observables. For the different temperatures and subsystems shown, the measured number statistics are in excellent agreement with microcanonical and canonical thermal ensembles, verifying the thermal character of the local density matrix. A grand-canonical ensemble reproduces the data very well as long as the subsystem is small compared to the full system. The error bars are the standard deviation of our observation over times between 10 and $20~\mathrm{ms}$. {\bf (D)} Thermalization occurs even for global quantities such as the full system interaction energy. The thermalization dynamics as calculated from our number-resolved images are in near agreement with exact numerical simulation and a canonical prediction.}
	\label{fig:ETH}
\end{figure*}

\section*{Local observables in the thermalized pure state}

Our comparisons between the entanglement entropy and thermal entropy suggest that the pure quantum state we study possesses thermalized properties. We can further examine the presence of thermalization by performing a series of measurements of local observables against which we compare the predictions of various thermal ensembles. As with the entanglement entropy, we also contrast our observations of the quenched thermalized state with the adiabatically prepared  ground state. In Figure~\ref{fig:local}A, we display the in-situ number density distribution on the six sites for the saturated quenched state and the (superfluid) ground state. While the ground state exhibits significant curvature, the quenched state exhibits a flat density distribution. This flat density distribution is consistent with a picture in which the constituents of the many-body system collectively thermalize, so that each site is in equilibrium with its neighbors and physically similar.

We can perform a more rigorous test of single-site thermalization by comparing the measured density matrix of each site with the reduced density matrix of a canonical thermal ensemble $\rho_A^T$ (Figure~\ref{fig:local}B). Our measurements of probabilities to observe a given particle number on a site completely characterize that single-site density matrix, because there are no coherences between different number states due to super-selection rules. With this measured density matrix, we can perform a quantitative comparison to a thermal ensemble using the trace distance ($\frac{1}{2}\mathrm{Tr}(\vert \rho_A^T -  \rho_A \vert)$) and quantum fidelity ($\mathrm{Tr}\left ( \sqrt{\sqrt{\rho_A^T} \rho_A \sqrt{\rho_A^T}}\right )$), both of which quantify the similarity of two mixed quantum states. After a short time, we see a quantum fidelity exceeding $99\%$ and a trace-distance that fluctuates between $0$ and $0.1$, indicating the similarity between the local density matrix of a verified pure state with the local density matrix of a thermal state. The correspondence between the observables of a pure state and thermal state depends on the equivalence of their reduced density matrices within the Hilbert space sampled by the observable. The measurement of Figure~\ref{fig:local}B therefore shows that observables for the single-site Hilbert space should agree with the predictions of thermal ensembles. 

We now focus on direct comparisons of observables with various thermal ensembles, and the theoretical justification for doing so. While we have focused on the role of entanglement entropy in producing thermal characteristics, the eigenstate distribution resulting from a quench (Figure~\ref{fig:ETH}A) determines the dynamics of observables, as well as their subsequent saturated values. It follows then that these populated eigenstates should clarify the origin of thermalization, which is the goal of ETH. The underlying explanation for ETH is that thermalizing, non-integrable systems possess excited eigenstates that look like nearly random vectors, or, equivalently, are described by a Hamiltonian that approximately conforms to random matrix theory~\cite{Deutsch1991, Alessio2015}. That is, for most bases, each eigenvector projects onto each basis vector with random quantum amplitude. The diffuse probability distribution of the eigenstates in most bases, such as the Fock state basis, is analogous to the chaotic dynamics of a closed classical mechanical system passing through every allowed point of phase space, and in the quantum case this has several consequences. Surprisingly, this chaotic assumption can be adapted to explain the saturation of measurement observables, the agreement of these saturated observables with thermal ensembles, and the presence of a volume law in the entanglement entropy~\cite{Page1993, Deutsch1991, KimThesis, Alessio2015}. And so, while in classical mechanical systems it is the chaos in the temporal dynamics that leads to entropy maximization and thermalization within thermodynamic constraints, in quantum thermalizing systems it is chaos in the energy eigenstates that generates the analogous behavior in the entanglement entropy,  and, in turn, thermalization. 

In Figure~\ref{fig:ETH}C,D, we compare our measurements to the predictions of thermal ensembles that are illustrated in Figure~\ref{fig:ETH}B. We also compare our results to a grand-canonical ensemble truncated to our total atom number~\cite{Supplement}: this ensemble perhaps most closely models how well the many-body state can act as a reservoir for its constituent subsystems. For each single-site and three-site observable, we show the atom number distributions for two different effective temperatures of $3.8J$ and $11J$, which are achieved by quenching to $J/U = 0.64$ and $J/U =2.6$, respectively.  The data is averaged in the saturated regime over times between $10$ and $20~\mathrm{ms}$, and the error bars are the standard deviation in the measured probabilities. The consistency within the the error bars indicates that in this temporal range our observations remain near the thermal predictions despite the presence of temporal fluctuations. For the single site subsystem, the data is in good agreement with all the ensembles considered. Despite the fact that the quenched state is in a large distribution of eigenstates, surprisingly, we find favorable agreement for the case of a single eigenstate ensemble: this illustrates a key principle of ETH, which holds that the reduced density matrix, and associated observables, vary slowly from eigenstate to eigenstate and are therefore relatively insensitive to breadth of the distribution of populated states from the quench. We perform the same comparisons to the three-site case in the bottom two panels. Here we also observe agreement with most ensembles, though, interestingly, there is relatively less agreement with the single eigenstate and grand-canonical ensembles, particularly for the lower temperature quench. This variation in agreement may suggest that these ensembles are more sensitive to the relative size of the traced out reservoir compared to the subsystem, which indicates directions of further experiments~\cite{Rigol2012, OlshaniiPC}.

The above measurements were on specific subsystems, but our measurements also allow extraction of the average global interaction energy (Figure~\ref{fig:ETH}D). Since the interaction term in Eq.~\ref{BHM} is diagonal in the Fock-state basis, we can use our measurements of the final particle configurations to compute the expectation value $\langle \hat{H}_{int} \rangle$. For the $T=3.8J$ data, we show a time scan indicating the initial growth in this quantity, which starts at zero since the initial state is a single particle per site.  These observations, at long times, are in near agreement with the canonical prediction. Interestingly, this measurement is sensitive to the entire six-site system as opposed to some subset of sites, which might suggest that it is global and unlikely to thermalize. Yet, $\langle \hat{H}_{int} \rangle$ undergoes thermalization because it is a sum of local operators, each of which thermalizes and is insensitive to the global purity of the full system.  The observed agreement is consistent with the idea that only a small set of operators, such as the global purity we measure or other specific fine-tuned state projectors, can truly distinguish the pure state we produce from a thermal state. 

\section*{Discussion}

Our observations speak to a natural translation between thermalizing quantum mechanical and classical mechanical systems composed of itinerant particles. Classical statistical mechanics relies on a fundamental assumption: a system in thermal equilibrium can be found in any microstate compatible with the thermodynamic constraints imposed on the system, and, as such, is described by an ensemble of maximal entropy~\cite{Ma, Huang}.  Although it is vastly successful, classical statistical mechanics does not itself justify this entropy maximization for closed systems~\cite{Ma, Alessio2015}, and an open systems approach only defers the question of thermalization to the union of the bath and system~\cite{Deutsch1991}. While ergodicity and time-averaging can provide a justification for entropy maximization in closed classical mechanical systems, ergodicity is not applicable on the same scale that statistical mechanics is successful, and time-averaging can require exponentially long times~\cite{Ma, Huang, Alessio2015}. The latter also obscures the fact that there is in reality only one system, which, nevertheless, is  well-modeled by an entropic ensemble~\cite{Ma}. Our studies, and beautiful recent theoretical work~\cite{Rigol2012, Deutsch2013,Grover2015}, hint towards a microscopic origin for entropy maximization in a single quantum state, namely that induced by the entanglement we measure. Quantum mechanics does not require time-averaging or thorny issues therein: a single quantum state obtains thermalized local observables, and these observables cannot distinguish this thermalized pure state from a mixed thermal ensemble of the same thermodynamic character. 

Our measurements open up several avenues for further investigation. Instead of operating with fixed total system size, we can study how thermalization and fluctuations depend on the size of the system considered~\cite{OlshaniiPC}. Conversely, studying integrable Hamiltonians where thermalization fails~\cite{Kinoshita2006}, and the structure of the associated eigenstate spectrum of such systems, could allow direct tests of the relationship between conserved quantities and thermalization of a quantum state. Lastly, applying these tools for characterizing the presence of thermalization and entanglement entropy could be powerful in studying many-body localization, where one of the key experimental signatures is the logarithmic growth of entanglement entropy at long times and suppression of precisely the thermalization we measure here~\cite{Prelov2008,Bardarson2012,Serbyn2013, Huse2015,Schreiber2015,Choi12016}. 

{\bf Acknowledgments}  We acknowledge helpful discussions with Soonwon Choi, Susannah Dickerson, Jens Eisert, Michael Foss-Feig, Daniel Greif, Matthew Headrick, David Huse, Maxim Olshanii, Cindy Regal, Johannes Schachenmayer, and Michael Wall. We are supported by grants from the NSF, Gordon and Betty Moore Foundations EPiQS Initiative (grant GBMF3795), an Air Force Office of Scientific Research MURI program, an Office of Naval Research MURI program, and an NSF Graduate Research Fellowship (M.R.).



\begin{thebibliography}{0}%
\makeatletter
\providecommand \@ifxundefined [1]{%
 \@ifx{#1\undefined}
}%
\providecommand \@ifnum [1]{%
 \ifnum #1\expandafter \@firstoftwo
 \else \expandafter \@secondoftwo
 \fi
}%
\providecommand \@ifx [1]{%
 \ifx #1\expandafter \@firstoftwo
 \else \expandafter \@secondoftwo
 \fi
}%
\providecommand \natexlab [1]{#1}%
\providecommand \enquote  [1]{``#1''}%
\providecommand \bibnamefont  [1]{#1}%
\providecommand \bibfnamefont [1]{#1}%
\providecommand \citenamefont [1]{#1}%
\providecommand \href@noop [0]{\@secondoftwo}%
\providecommand \href [0]{\begingroup \@sanitize@url \@href}%
\providecommand \@href[1]{\@@startlink{#1}\@@href}%
\providecommand \@@href[1]{\endgroup#1\@@endlink}%
\providecommand \@sanitize@url [0]{\catcode `\\12\catcode `\$12\catcode
  `\&12\catcode `\#12\catcode `\^12\catcode `\_12\catcode `\%12\relax}%
\providecommand \@@startlink[1]{}%
\providecommand \@@endlink[0]{}%
\providecommand \url  [0]{\begingroup\@sanitize@url \@url }%
\providecommand \@url [1]{\endgroup\@href {#1}{\urlprefix }}%
\providecommand \urlprefix  [0]{URL }%
\providecommand \Eprint [0]{\href }%
\providecommand \doibase [0]{http://dx.doi.org/}%
\providecommand \selectlanguage [0]{\@gobble}%
\providecommand \bibinfo  [0]{\@secondoftwo}%
\providecommand \bibfield  [0]{\@secondoftwo}%
\providecommand \translation [1]{[#1]}%
\providecommand \BibitemOpen [0]{}%
\providecommand \bibitemStop [0]{}%
\providecommand \bibitemNoStop [0]{.\EOS\space}%
\providecommand \EOS [0]{\spacefactor3000\relax}%
\providecommand \BibitemShut  [1]{\csname bibitem#1\endcsname}%
\let\auto@bib@innerbib\@empty
\end{thebibliography}%


\begin{thebibliography}{10}

\bibitem{Sakurai}
J.~J. Sakurai, {\em {Modern Quantum Mechanics}} (Addison Wesley Longman, 1993).

\bibitem{CardyOneD}
P. Calabrese and J. Cardy, {\em Evolution of entanglement entropy in
  one-dimensional systems}, Journal of Statistical Mechanics: Theory and
  Experiment {\bf 2005},  P04010  (2005).

\bibitem{Amico2008}
L. Amico, R. Fazio, A. Osterloh, and V. Vedral, {\em Entanglement in many-body
  systems}, Rev. Mod. Phys. {\bf 80},  517  (2008).

\bibitem{Daley2012}
A.~J. Daley, H. Pichler, J. Schachenmayer, and P. Zoller, {\em Measuring
  Entanglement Growth in Quench Dynamics of Bosons in an Optical Lattice},
  Phys. Rev. Lett. {\bf 109},  020505  (2012).

\bibitem{Schachenmayer2013}
J. Schachenmayer, B.~P. Lanyon, C.~F. Roos, and A.~J. Daley, {\em Entanglement
  Growth in Quench Dynamics with Variable Range Interactions}, Phys. Rev. X
  {\bf 3},  031015  (2013).

\bibitem{Deutsch1991}
J.~M. Deutsch, {\em Quantum statistical mechanics in a closed system}, Phys.
  Rev. A {\bf 43},  2046  (1991).

\bibitem{Olshanii2008}
M. Rigol, V. Dunjko, and M. Olshanii, {\em Thermalization and its mechanism for
  generic isolated quantum systems}, Nature {\bf 452},  854  (2008).

\bibitem{Eisert2015}
J. Eisert, M. Friesdorf, and C. Gogolin, {\em Quantum many-body systems out of
  equilibrium}, Nat Phys {\bf 11},  124  (2015).

\bibitem{Shankar1985}
R.~V. Jensen and R. Shankar, {\em Statistical Behavior in Deterministic Quantum
  Systems with Few Degrees of Freedom}, Phys. Rev. Lett. {\bf 54},  1879
  (1985).

\bibitem{SrendickiETH}
M. Srednicki, {\em Chaos and quantum thermalization}, Phys. Rev. E {\bf 50},
  888  (1994).

\bibitem{Rigol2012}
L.~F. Santos, A. Polkovnikov, and M. Rigol, {\em Weak and strong typicality in
  quantum systems}, Phys. Rev. E {\bf 86},  010102  (2012).

\bibitem{Deutsch2013}
J.~M. Deutsch, H. Li, and A. Sharma, {\em Microscopic origin of thermodynamic
  entropy in isolated systems}, Phys. Rev. E {\bf 87},  042135  (2013).

\bibitem{Alessio2015}
L. D'Alessio, Y. Kafri, A. Polkovnikov, and M. Rigol, {\em From Quantum Chaos
  and Eigenstate Thermalization to Statistical Mechanics and Thermodynamics},
  arXiv:1509.06411v1  (2015).

\bibitem{Martinis2016}
C. Neill {\it et~al.}, {\em Ergodic dynamics and thermalization in an isolated
  quantum system}, arXiv:1601.00600  (2016).

\bibitem{Schaetz2015}
G. Clos, D. Porras, U. Warring, and T. Schaetz, {\em Time-resolved observation
  of thermalization in an isolated quantum system}, arXiv:1509.07712  (2015).

\bibitem{Trotzky2012}
S. Trotzky {\it et~al.}, {\em Probing the relaxation towards equilibrium in an
  isolated strongly correlated one-dimensional Bose gas}, Nat Phys {\bf 8},
  325  (2012).

\bibitem{Langen2013}
T. Langen, R. Geiger, M. Kuhnert, B. Rauer, and J. Schmiedmayer, {\em Local
  emergence of thermal correlations in an isolated quantum many-body system},
  Nat Phys {\bf 9},  640  (2013).

\bibitem{Geiger2014}
R. Geiger, T. Langen, I.~E. Mazets, and J. Schmiedmayer, {\em Local relaxation
  and light-cone-like propagation of correlations in a trapped one-dimensional
  Bose gas}, New Journal of Physics {\bf 16},  053034  (2014).

\bibitem{Langen2015}
T. Langen {\it et~al.}, {\em Experimental observation of a generalized Gibbs
  ensemble}, Science {\bf 348},  207  (2015).

\bibitem{Huse2015}
R. Nandkishore and D.~A. Huse, {\em Many-Body Localization and Thermalization
  in Quantum Statistical Mechanics}, Annual Review of Condensed Matter Physics
  {\bf 6},  15  (2015).

\bibitem{GreinerQGM}
W.~S. Bakr {\it et~al.}, {\em Probing the Superfluid-to-Mott Insulator
  Transition at the Single-Atom Level}, Science {\bf 329},  547  (2010).

\bibitem{BlochQGM}
J.~F. Sherson {\it et~al.}, {\em Single-atom-resolved fluorescence imaging of
  an atomic Mott insulator}, Nature {\bf 467},  68  (2010).

\bibitem{Zupancic2016}
P. Zupancic {\it et~al.}, {\em Ultra-precise holographic beam shaping for
  microscopic quantum control}, arXiv:1604.07653  (2016).

\bibitem{Supplement}
See supplementary material.

\bibitem{Islam2015}
R. Islam {\it et~al.}, {\em Measuring entanglement entropy in a quantum
  many-body system}, Nature {\bf 528},  77  (2015).

\bibitem{Sackett2000}
C.~A. Sackett {\it et~al.}, {\em Experimental entanglement of four particles},
  Nature {\bf 404},  256  (2000).

\bibitem{JakschPRA}
R.~N. Palmer, C. Moura~Alves, and D. Jaksch, {\em Detection and
  characterization of multipartite entanglement in optical lattices}, Phys.
  Rev. A {\bf 72},  042335  (2005).

\bibitem{Hazzard2014}
K.~R.~A. Hazzard {\it et~al.}, {\em Quantum correlations and entanglement in
  far-from-equilibrium spin systems}, Phys. Rev. A {\bf 90},  063622  (2014).

\bibitem{HHHH}
R. Horodecki, P. Horodecki, M. Horodecki, and K. Horodecki, {\em Quantum
  entanglement}, Rev. Mod. Phys. {\bf 81},  865  (2009).

\bibitem{Horodecki1996a}
R. Horodecki and M. Horodecki, {\em Information-theoretic aspects of
  inseparability of mixed states}, Phys. Rev. A {\bf 54},  1838  (1996).

\bibitem{lightcone}
M. Cheneau {\it et~al.}, {\em Light-cone-like spreading of correlations in a
  quantum many-body system}, Nature {\bf 481},  484  (2012).

\bibitem{Richerme2014}
P. Richerme {\it et~al.}, {\em Non-local propagation of correlations in quantum
  systems with long-range interactions}, Nature {\bf 511},  198  (2014).

\bibitem{CardyGS}
P. Calabrese and J. Cardy, {\em Entanglement entropy and quantum field theory},
  Journal of Statistical Mechanics: Theory and Experiment {\bf 2004},  P06002
  (2004).

\bibitem{Plenio2010}
J. Eisert, M. Cramer, and M.~B. Plenio, {\em Colloquium: Area laws for the
  entanglement entropy}, Rev. Mod. Phys. {\bf 82},  277  (2010).

\bibitem{Grover2015}
J.~R. Garrison and T. Grover, {\em Does a single eigenstate encode the full
  Hamiltonian?}, arXiv:1503.00729  (2015).

\bibitem{Grover2014}
T. Grover and M.~P.~A. Fisher, {\em Entanglement and the sign structure of
  quantum states}, Phys. Rev. A {\bf 92},  042308  (2015).

\bibitem{Wolf2008}
M.~M. Wolf, F. Verstraete, M.~B. Hastings, and J.~I. Cirac, {\em Area Laws in
  Quantum Systems: Mutual Information and Correlations}, Phys. Rev. Lett. {\bf
  100},  070502  (2008).

\bibitem{Page1993}
D.~N. Page, {\em Average entropy of a subsystem}, Phys. Rev. Lett. {\bf 71},
  1291  (1993).

\bibitem{KimThesis}
K. Hyungwon, , Quantum Nonequilibrium Dynamics: Transport, Entanglement, and
  Thermalization, PhD thesis  (2014).

\bibitem{OlshaniiPC}
\emph{Private communication} Maxim Olshanii.

\bibitem{Ma}
S.~K. Ma, {\em {Statistical Mechanics}} (World Scientific, Singapore, 1985).

\bibitem{Huang}
K. Huang, {\em {Statistical Mechanics}} (John Wiley and Sons, Inc., ADDRESS,
  1963).

\bibitem{Kinoshita2006}
T. Kinoshita, T. Wenger, and D.~S. Weiss, {\em A quantum Newton's cradle},
  Nature {\bf 440},  900  (2006).

\bibitem{Prelov2008}
M. \ifmmode \check{Z}\else \v{Z}\fi{}nidari\ifmmode~\check{c}\else \v{c}\fi{},
  T.~c.~v. Prosen, and P. Prelov\ifmmode~\check{s}\else \v{s}\fi{}ek, {\em
  Many-body localization in the Heisenberg $XXZ$ magnet in a random field},
  Phys. Rev. B {\bf 77},  064426  (2008).

\bibitem{Bardarson2012}
J.~H. Bardarson, F. Pollmann, and J.~E. Moore, {\em Unbounded Growth of
  Entanglement in Models of Many-Body Localization}, Phys. Rev. Lett. {\bf
  109},  017202  (2012).

\bibitem{Serbyn2013}
M. Serbyn, Z. Papi\ifmmode~\acute{c}\else \'{c}\fi{}, and D.~A. Abanin, {\em
  Universal Slow Growth of Entanglement in Interacting Strongly Disordered
  Systems}, Phys. Rev. Lett. {\bf 110},  260601  (2013).

\bibitem{Schreiber2015}
M. Schreiber {\it et~al.}, {\em Observation of many-body localization of
  interacting fermions in a quasirandom optical lattice}, Science {\bf 349},
  842  (2015).

\bibitem{Choi12016}
J.-y. Choi {\it et~al.}, {\em Exploring the many-body localization transition
  in two dimensions}, Science {\bf 352},  1547  (2016).

\end{thebibliography}

\clearpage

\begin{widetext}


\section*{Supplementary material}

Adam M. Kaufman, M. Eric Tai, Alexander Lukin, Matthew Rispoli,
Robert Schittko, Philipp M. Preiss, Markus Greiner

Department of Physics, Harvard University, Cambridge, Massachusetts 02138, USA

*To whom correspondence should be addressed. E-mail: greiner@physics.harvard.edu.

This PDF includes:

- Supplementary text
- Figures S1-S3
- Tables S1-S2

# 1 Eigenstate Thermalization Hypothesis

Here we provide a brief summary of the Eigenstate Thermalization Hypothesis (ETH) and its relationship to our experiment. Generically, a quenched quantum state consists of a superposition of many-body energy eigenstates, each of which evolves according to the frequency of the associated eigenenergy $E_n$. For the state $|\psi(t)\rangle = \sum_n c_n e^{-iE_n t} |n\rangle$, the observable $\langle \mathcal{O}(t) \rangle$ evolves according to,

$$\langle \mathcal{O}(t) \rangle = \sum_{\alpha,\beta} c_\alpha^* c_\beta e^{i(E_\alpha - E_\beta)t/\hbar} \mathcal{O}_{\alpha\beta}$$

$$\langle \mathcal{O}(t) \rangle = \underbrace{\sum_{\alpha} |c_\alpha|^2 \mathcal{O}_{\alpha\alpha}}_{\mathcal{S}_{\text{diag}}} + \underbrace{\sum_{\alpha,\beta \neq \alpha} c_\alpha^* c_\beta e^{i(E_\alpha - E_\beta)t/\hbar} \mathcal{O}_{\alpha\beta}}_{\mathcal{S}_{\text{off}}}$$

where $\mathcal{O}_{\alpha\beta} = \langle \alpha | \mathcal{O} | \beta \rangle$ for the energy eigenstates $|\alpha\rangle$ and $|\beta\rangle$. We consider the system to thermalize if

1. the first term, $\mathcal{S}_{\text{diag}}$, takes a value $\overline{\mathcal{O}}$ that matches the microcanonical prediction
2. the second term, $\mathcal{S}_{\text{off}}$, has only small fluctuations around zero for long times

Regarding condition 1, the microcanonical ensemble predicts a value for macroscopic observables that depends only on the average energy of the system. However, $\mathcal{S}_{\text{diag}}$ explicitly contains terms related to the initial population distribution, which suggests the saturated value of the observable depends on initial details of the system, rather than just the average energy. ETH resolves this puzzle by stating that a quantum quench populates mostly the eigenstates far from the edge of the spectrum, and that these eigenstates approximate those of random matrix theory. ETH proposes that these eigenstates look locally thermal and the expectation value of local observables varies smoothly between eigenstates close in energy. This implies that the exact probabilities from condition $\mathcal{S}_{\text{diag}}$ are not quantitatively relevant in the sum, since the $\mathcal{O}_{\alpha,\alpha}$ can be approximately factored out. Regarding condition 2, while for times short compared to the spread in populated eigenfrequencies the relative phases of each component in $\mathcal{S}_{\text{off}}$ are fine-tuned to match the initial state, at long times the relative phases are randomized. Ostensibly, this dephasing ensures that $\mathcal{S}_{\text{off}} \to 0$, but the time scale for this process is given by the "typical" value of the smallest gap in the spectrum. In a general many-body system, this value can be exponentially small which would lead to an infinite thermalization time, but for the systems where ETH applies level repulsion ensures that this "typical" gap value stays finite (13). As important, ETH states that the off-diagonals of $\mathcal{O}$ in the eigenbasis are negligible compared to the on-diagonals, so that this second term damps to a value which does not influence the steady-state of $\langle \mathcal{O} \rangle$. For a more detailed summary of ETH, we point the reader to Refs. (7, 13).

# 2 Computing expectation values in thermalized systems

ETH implies an equivalence between the local expectation values of a quenched many-body state and those of the thermal density matrix with the same average total energy as the many-body state. For the reported experiments, our system is initialized into the ground state, $|\psi_0\rangle$, of an initial Hamiltonian, $\mathcal{H}_0$, in the atomic limit. At $t=0$, we quench the system into a Hamiltonian, $\mathcal{H}_q$, after which the system is allowed to evolve for a variable amount of time. Comparing to the data at longs times ($10-20$ ms, where we observe saturation), we can compute predictions for the expectation values of various local observables based upon different thermodynamic ensembles. These predictions are computed using the following procedures.

### Microcanonical Ensemble

The microcanonical ensemble is an equal probability statistical mixture of all the eigenstates that lie within an energy interval given by the initial state $|\psi_0\rangle$. In the quenched Hamiltonian, the initial state has an energy

$$E^{(0)} \equiv \langle \psi_0 | \mathcal{H}_q | \psi_0 \rangle$$

while the eigenstates of $\mathcal{H}_q$, $\left|\phi_i^{(q)}\right\rangle$, have energies $E_i^{(q)}$. The microcanonical ensemble is then composed of the $N_{\text{MC}}$ number of eigenstates for which $|E_i^{(q)} - E^{(0)}| < \delta E$. For our numerical data, we have chosen $\delta E = 0.2J$, but the ensemble predictions are insensitive to the precise value of $\delta E$. The microcanonical ensemble can be represented by the thermal density matrix

$$\rho_{ij}^{MC} = \begin{cases} \frac{1}{N_{\text{MC}}}, & \text{if } i = j \text{ and } |E_i^{(q)} - E^{(0)}| < \delta E \\ 0, & \text{else} \end{cases}.$$

## Canonical Ensemble

The canonical ensemble is a statistical mixture of all the eigenstates in the system weighted by each state's Boltzmann factor, $\exp(-E_i^{(q)}/k_B T)$. The temperature in the Boltzmann factor is fixed through the stipulation that the average energy of this thermal ensemble matches the energy of the initial state, i.e. we choose $T$ such that

$$\text{Tr}(\mathcal{H}_q \rho^{CE}) = \langle \psi_0 | \mathcal{H}_q | \psi_0 \rangle,$$

where the thermal density matrix $\rho^{CE}$ has the following construction,

$$\rho_{ij}^{CE} = \begin{cases} e^{-\frac{E_i^{(q)}}{k_B T}}, & \text{if } i = j \\ 0, & \text{else} \end{cases}.$$

## Single Eigenstate Ensemble

Energy eigenstates of systems conforming to ETH are surmised to appear thermal in local observables. We numerically calculate the eigenstates of the quenched Hamiltonian and compare the experimentally observed local counting statistics to the prediction from the single full system eigenstate $\left|\phi_i^{(q)}\right\rangle$ that is closest in energy to the expectation value $E^{(0)}$. The expectation value in this case is given by,

$$\langle \mathcal{A} \rangle_{\text{SE}} = \left\langle \phi_i^{(q)} \right| \mathcal{A} \left| \phi_i^{(q)} \right\rangle.$$

## Diagonal Ensemble

The diagonal ensemble is a statistical mixture of all eigenstates of the full Hamiltonian $\mathcal{H}_q$, with the weights given by their amplitudes after quench.

$$\rho_{ij}^D = \begin{cases} |\langle \psi_0 | \phi_i^{(q)} \rangle|^2, & \text{if } i = j \\ 0, & \text{else} \end{cases}.$$

It carries all information about the amplitudes of the eigenstates but ignores all their relative phases.

## Grand Canonical Ensemble

The grand-canonical ensemble requires calculating the temperature and chemical potential for the subsystem associated to the observable of interest. For example, the top (bottom) row of FIG. 6C pertains to the subsystem consisting of the third site (the first three sites) of the chain. We calculate the temperature and chemical potential for the subsystem as follows. Because the energy and particle number within the subsystem are not conserved during the quench dynamics, we must compute the average energy $\langle E_A \rangle$ and average number $\langle N_A \rangle$ within the subsystem numerically. We time-evolve the full many-body state to the thermalized regime, then compute the reduced density matrix for the subsystem, with which we can calculate $\langle N_A \rangle$ and $\langle E_A \rangle$. We note that the average energy of nearly all the subsystems is very close to that of the full system (zero), while the average number is nearly consistent with unity particle density. For the single site subsystems, however, there is no tunneling term to offset the interaction energy, and therefore these subsystems have non-zero energy. We perform this full calculation to account for finite-size effects that cause

small temporal energy and number fluctuations. If we neglect the energy fluctuations, the grand-canonical predictions (described below) are negligibly different.

After the above calculations, we can compute the chemical potential and temperature. Using each $H_A^N$, the subsystem Bose-Hubbard Hamiltonian with $N$ particles, we compute the eigenstates ($|E_A^{N,i}\rangle$, where $i$ indexes the eigenstate) and energies ($E_A^{N,i}$) for each particle sector. We seek $T$ and $\mu$ such that,

$$\langle N_A \rangle = \langle N_{\text{GCE}} \rangle = \frac{1}{Z} \sum_{i,N} N e^{-(E_A^{N,i} - \mu N)/k_B T}$$

and,

$$\langle E_A \rangle = \langle E_{\text{GCE}} \rangle = \frac{1}{Z} \sum_{i,N} E_A^{N,i} e^{-(E_A^{N,i} - \mu N)/k_B T},$$

where for each particle number $N$, the index $i$ is summed over all eigenstates within that number sector. The partition function, $Z$, is the overall normalization. These equations are numerically solved to find $\mu$ and $T$. With these in hand, we arrive at the grand-canonical ensemble,

$$\rho_{\text{GCE}} = \frac{1}{Z} \sum_{i,N} |E_A^{N,i}\rangle\langle E_A^{N,i}| e^{-(E_A^{N,i} - \mu N)/k_B T}.$$

### Observables

For all statistical ensembles above, expectation values of an observable $\mathcal{A}$ are calculated from the density matrix as

$$\langle \mathcal{A} \rangle = \text{Tr}(\mathcal{A}\rho),$$

where $\rho$ is the full system density matrix corresponding to the appropriate ensemble.

## 3 Entropies

### 3.1 Rényi Entropy

All entanglement entropy values discussed in the main text are defined as a function of the reduced density matrix of the subsystem. This applies also for the thermal entropies quoted in the main text, where we use the reduced density matrix of the canonical ensemble described above. The entropy metric used for comparison between the quantum system we measure and the canonical thermal ensemble is the Rényi entropy. In general, the n-th order Rényi entropy is defined for a reduced density matrix of subsystem A as

$$S_n(\rho_A) = \frac{1}{1-n} \log\left(\text{Tr}[\rho_A^n]\right),$$

where the reduced density matrix $\rho_A$ is defined by tracing out all degrees of freedom of the system that do not include subsystem A

$$\rho_A = \text{Tr}_B(\rho_{AB}).$$

Experimentally, we interfere two identical quantum states to obtain simultaneously the global and local purity ($\text{Tr}(\rho_A^2)$). Therefore the relevant order Rényi entropy is the second-order ($n = 2$) Rényi entropy which is defined for a density matrix of subsystem as

$$S_2(\rho_A) = -\log\left(\text{Tr}[\rho_A^2]\right),$$

and is also important qualitatively as a lower bound of the von Neumann entropy. When discussing the global purity as an entropy, we use the Rényi formulation in terms of the global density matrix $\rho$ as opposed to a reduced density matrix $\rho_A$. Lastly, note that we use logarithms to base $e$ throughout.

It is important to stress that the thermodynamic relations defined from statistical mechanics with the von Neumann definition do not directly apply for the Rényi definition. However, both quantities measure the incoherent diffusion in Hilbert space associated with entropy, and the Rényi entropy is directly accessible by our measurements.

## 3.2 Simulations

All simulations use measured experimental parameters from Table S1 to construct the Hamiltonian. The shown theory curves are computed through direct diagonalization of the Hamiltonian given the relevant Bose-Hubbard parameters. The inital state is the exact Mott insulating ground state of both lattices at $45E_r$. The 1-D quench dynamics are then simulated by projecting this initial state onto the eigenstates of the quenched Bose-Hubbard Hamiltonian and evolved in time by their corresponding eigenvalues. The full density matrix can then be constructed at any time $t$ and traced for either the on-site number statistics or entanglement entropy to be compared with the data. The plotted theory have zero-fit parameters, and are only corrected for overall offsets due to the measured residual extensive entropy in the data (see Section 4). We also include a temporal offset of 0.5 ms that is the same for all theory calculations, which is based on an independent numerical model of the (nearly perfectly diabatic) ramp that realizes the quench. Besides this offset, we have confirmed numerically that the degree to which this ramp is not perfectly diabatic does not influence the results.

# 4  Experimental Sequence

### State Preparation

Our experiments start with a unity filling, two-dimensional Mott insulator of $^{87}$Rb atoms in a deep lattice ($V_x = V_y = 45E_r$) with 680 nm spacing. We utilize a DMD in the Fourier plane to initialize a plaquette of $2\times 6$ atoms using a procedure outlined in previous work (24). We achieve a single site plaquette loading fidelity of $\sim 93\%$, which is limited by the fidelity of the initial Mott insulator and losses during the preparation sequence.

To study the dynamics in the quenched system, we project an optical potential consisting of two narrow Gaussian peaks separated by 6 lattice sites in the $x$-direction and a flat-top profile in the $y$-direction. This confinement provides the atoms with a "box"-like potential superimposed upon and registered to the lattice position. By changing $V_x$ from $45E_r$ to $6E_r$ in 0.75 ms (70 ms) while retaining $V_y = 45E_r$ we realize identical diabatic (adiabatic) evolution in 1-D 6-site chains. After some evolution time we rapidly freeze the dynamics by ramping the lattice along $x$-direction to $20E_r$. We can then conduct two types of experiments on the final state of the system: perform a single site resolved atom number counting or measuring the entropy of the state by means of a beam splitter operation.

### Single-site resolved atom number counting

In this section, we outline our procedure for counting the total atom number on each site of the 2x6 plaquette. After freezing the atom dynamics, we project a narrow (along the $y$-direction) Gaussian potential. We then drop $V_y = 0E_r$ while keeping $V_x = 45E_r$, separating the atoms away from the partition while preserving their position in the $x$-direction. After 2 ms of free evolution in the half tubes we pin the atom positions with the imaging lattice ($3000E_r$) and perform fluorescence imaging of the atoms. The total population of each half tube corresponds to the original population of a particular lattice site in each of the two copies. During the expansion process atoms delocalize uniformly over $\sim$50 lattice sites. The majority of our outcomes ($\sim 96\%$) contain 3 or fewer particles per site which results in probability of particle loss due to parity projection $< 6\%$. We do not apply a correction for this effect because it is smaller than the statistical error for the majority of data points.

### Entanglement entropy

We measure global purity and entanglement entropy by applying an atomic beam splitter and interfering the two copies of the 6-site system. We perform the beam splitter operation by projecting a double-well potential along $y$ that is superimposed on the lattice. The tunnel-coupling of the double-well is set by a combination of both the $y$ lattice and this projected potential, which is nearly uniform along the $x$ direction. After diabatically reducing $V_y = 2E_r$, the atoms tunnel in the double-wells for 0.33 ms, which corresponds to the single particle balanced beam splitter time (24). This operation results in a two-particle oscillation

contrast of ∼ 96%. We infer the purity of any given subsystem by computing the average parity given by $\text{Tr}(\rho_A^2) = \langle P^i \rangle = \langle \prod_{k \in A} p_k^i \rangle$, where $i = 1, 2$ denotes one of the output modes of the double-well and $p_k^i = \pm 1$ denotes the parity of a given mode (4).

### Extensive background entropy

The imperfect fidelity of the beam splitter operation reduces the interference contrast between the two many-body systems. The measured purities hence underestimate the purity of the many-body states produced in the experiment.

We verify experimentally that this entropy background contributed by imperfections is extensive. For a separable many-body state, such as a Mott insulator in the atomic limit, we observe an entropy of 0.34 for the full system, or 0.06 per site.

For the relevant case of superfluid ground states and highly excited quenched states, the measured full system entropy is increased to 0.63. We attribute this increase in measured entropy to the sensitivity of such states to differences in the Hamiltonian between the two copies: During the short 3 ms hold time between the state preparation and the beam splitter operation, potential differences on the 5 Hz level can lead to a differential dephasing between the two copies and a reduced interference contrast. We therefore minimize the hold time and use an intermediate lattice depth ($20 E_r$).

We confirm the additive, extensive nature of this background entropy by subtracting the theoretical predicted value of entropy from the measured one as a function of the system size, which shows linear growth of this quantity within our statistics (see Fig. S2). By fitting the data we extract the slope of the curve and apply the correction given by the subsystem size to different plots (see Table S2). With the exception of very short times, these data affirm that the extensive corrections are substantially smaller than the entanglement entropy we measure.

### Bose-Hubbard parameters

Experimental parameters for the state preparation and quenches to different temperatures are shown in Table S1. The temperatures tabulated correspond to the canonical ensemble, as described in Section 2. We calibrate lattice depths through modulation spectroscopy in a deep $45 E_r$ lattice with a typical uncertainty of $\pm 2\%$ and obtain $J$ from a band structure calculation. $U$ is measured through photon-assisted tunneling at a lattice depth of $V_x = 16 E_r$ and numerically extrapolated to lower depths, taking into account corrections due to higher bands. Throughout the manuscript, the measured entanglement entropy and counting statistics are compared to numerical solutions of the Schrödinger equation with the appropriate Bose-Hubbard parameters.

## 5 Data analysis

### Post-selection

Before analyzing the data, we post-select on the outcomes containing 6 particles in each copy for the site-resolved number detection (FIG. 5,6). We post-select on total of 12 atoms for in both copies for experiments in which we perform the beam splitter operation. Outcomes containing lower particle number are attributed to the instances of imperfect plaquette loading or atom loss during the experimental sequence. We thus retain approximately 30% of the original data.

### Data averaging

For all entanglement entropy data we average over both copies after the beam splitter operation. Additionally, we take the mean of equivalent symmetric subsystems in each copy for entanglement entropy dynamics (FIG. 3), for the volume law data (FIG. 4A) we average over all contiguous subsystems of the given size. Finally, for the data in FIG. 4B,C, we average over all contiguous and non-contiguous subsystems of the given volume, because the mutual information can sample correlations between non-contiguous subsystems. For all of the entanglement entropy data, the error bars are S.E.M. For all of the number counting, the error

bars are S.E.M. except when stated otherwise. For the entanglement entropy and number counting data, typically 100-200 post-selected runs of the experiment are used to create each data point in the plots.

## Quantifying the early time time dynamics of the entanglement entropy

In an inset of FIG. 3, we quantify the growth of entanglement entropy in the early time dynamics. We approximate this growth as linear, while the long time dynamics as flat. We fit the data with a piecewise function,

$$S_2(t) = \begin{cases} S_g \cdot t, & \text{if } t \leq t_{\text{sat.}} \\ S_g \cdot t_{\text{sat.}}, & \text{else} \end{cases},$$

where the slope $S_g$ and saturation time $t_{\text{sat.}}$ are left as free parameters (Fig. S3). The fitted value of $S_g$ is quoted in the text for the slope.

# Figures and Tables

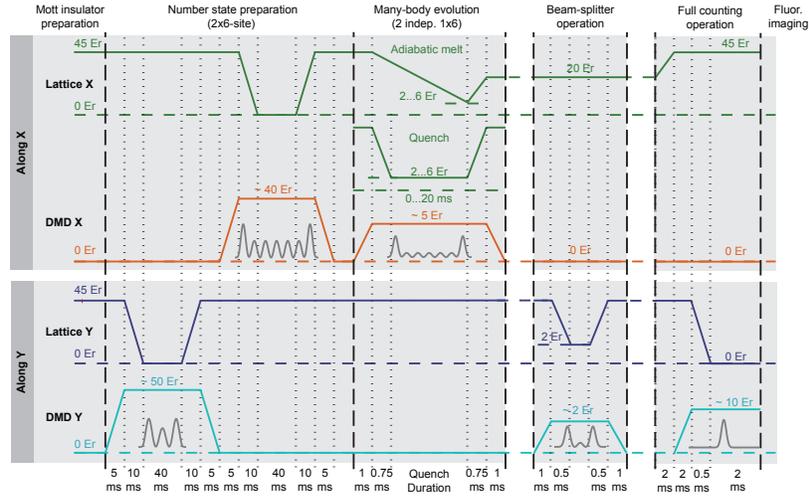

Figure S1: Schematic showing the ramps of the lattices and DMD potentials in the $x$- and $y$-directions during the experimental sequence. The projected DMD potentials are drawn in the labeled direction where the orthogonal direction of the potential is a smooth, flat-top profile. The break for the beam splitter operation is shown as an optional step taken during the sequence. Performing this operation results in the sequence measuring the local and global system purity while not performing this step results in the sequence measuring the on-site number statistics. All ramps are exponential in depth as a function of time.

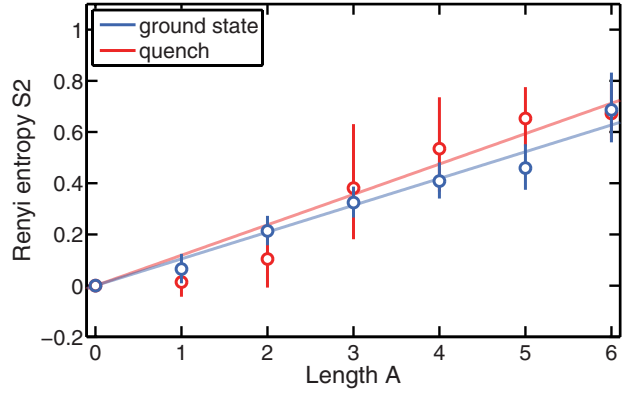

Figure S2: The difference between the measured Rényi entanglement entropy and theoretical calculation as a function of subsystem size. We use the entanglement entropy data of FIG. 4, averaged over contiguous and non-contiguous sub-systems; the subtracted theory exhibits the same averaging. The linear scaling indicates the presence of an extensive residual entropy background that we attribute to imperfections in the measurement protocol.

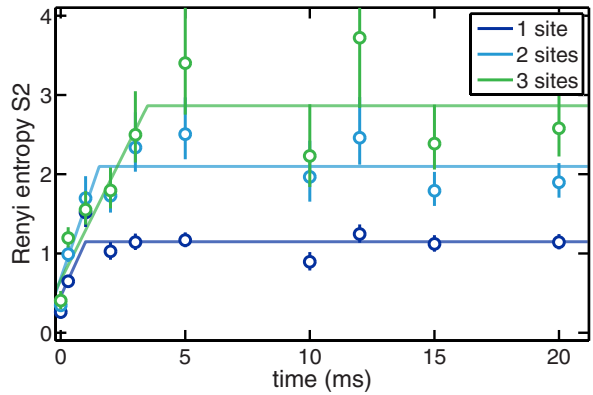

Figure S3: Piecewise linear fits to entanglement entropy dynamics for the three data sets of FIG. 3.

|  | $V_x$ [$E_r$] | $V_y$ [$E_r$] | $J_x/(2\pi)$ [Hz] | $U/(2\pi)$ [Hz] | $T/(2\pi)$ [Hz] |
|---|---|---|---|---|---|
| initial state | 45 | 45 | 0.07 | 172 | 0 |
| quench I | 6 | 45 | 66 | 103 | 249 |
| quench II | 2 | 45 | 178 | 68 | 1965 |

Table S1: Experimental parameters for the Hubbard chains. The initial Mott insulating state deep in the atomic limit $J_x \ll U$ is projected onto Bose-Hubbard chains with larger values of $J_x/U$, corresponding to different effective canonical temperatures $T$. All numerical simulations in the main text use the parameter values listed here.

|  | Theory | Data |
|---|---|---|
| Figure 2B | N/A | No Corrections |
| Figure 3 | Offset Added | No Corrections |
| Figure 4 | No Corrections | Extensive Entropy Subtracted |

Table S2: Listing of all figures that contain data and the numerical corrections applied based upon residual extensive entropy in the system due to beam splitter infidelity.


\end{widetext}

\newpage

\end{document}